\documentstyle[preprint,eqsecnum,aps]{revtex}
\begin{document}
\draft
\title{Surface superconductivity and $H_{c3}$ in UPt$_3$}
\author{Ding Chuan Chen and Anupam Garg}
\address{Department of Physics and Astronomy and Materials Research Center,\\
Northwestern University, Evanston, Illinois 60208
}
\date{\today}
\maketitle
\begin{abstract}
Surface superconductivity is studied within Ginzburg-Landau theory for
two classes of models for the order parameter of UPt$_{3}$. The first
class assumes two independent one-dimensional order parameters ($AB$
models), while the second assumes a single two-dimensional order
parameter ($E$ models). $H_{c3}$ is calculated for all cases where the
surface normal and magnetic field lie along high symmetry directions.
Assuming specular reflection, it is found that except when ${\bf
H}\parallel{\bf\hat c}$, the ratio $H_{c3}/H_{c2}$ is either unity or equals
its `s-wave' value 1.695, although the precise $H_{c3}$ vs. $T$ curve
predicted by the $AB$ and $E$ models differs for various geometries.
The results are compared with recent experiments, and predictions are
made for future experiments.
\end{abstract}


\section{Introduction}
UPt$_3$ is generally believed to be a reduced symmetry superconductor,
i.e., one that  breaks rotational and/or time reversal symmetries of
the normal state, in addition to the U(1) gauge
symmetry\cite{refSauls,refTail}. The strongest basis for this belief is
the phase diagram which displays  multiple superconducting
phases\cite{refPDa,refPDb,refPDc,refPDd,refPDe,refPDf}, just as liquid
$^3$He has multiple superfluid phases. In this paper we shall study
surface superconductivity in UPt$_3$, and see if competing models make
differing predictions which might be experimentally tested.

We briefly review the existing models for the order parameter in
UPt$_3$. There are two main classes. The first class, which we shall
refer to as `$E$'
models\cite{refSauls,refEma,refEmb,refEmc,refEmd,refEme}, entails a
single two-component order parameter, which may belong to any of the
two-dimensional representations ($E_{1g}$, $E_{2g}$, $E_{1u}$,
$E_{2u}$) of $D'_{6h}$, the point group of normal UPt$_3$\cite{refa}.
The second class of models posits two independent order parameters
belonging to different
representations\cite{refEmd,refCGprl,refGCprb,refMin,refHeid}.
In one subclass \cite{refb}, which can be analyzed in great detail,
comprising what
we shall call `$AB$' models, both order parameters have the same
parity; one transforms as an $A$ representation and the other as a $B$
representation of $D'_{6h}$.  Thus, if the common parity is
even\cite{refc}, the pair of order parameters transforms in one of four
possible ways: ($A_{1g}$, $B_{1g}$), ($A_{1g}$, $B_{2g}$), ($A_{2g}$,
$B_{1g}$), ($A_{2g}$, $B_{2g}$).  This division into two classes is
useful because in the simplest form of Ginzburg-Landau (GL) theory in
which only terms that are formally of order $(1-T/T_c)^2$ are kept, all
$E$ models have the same formal GL free energy, and all $AB$ models
have the same GL free energy. This leads one to sometimes refer to them
in the singular as the $E$ model or the $AB$ model.

The $E$ model invokes a symmetry breaking field (SBF), usually taken to
be a weak antiferromagnetism that sets in at 5~K\cite{refaepp,refGold},
in order to split the superconducting transition in zero field into two
transitions as observed. The $AB$ model posits two nearby transition
temperatures by fiat.  The original version of the $E$ model was shown
to be incompatible \cite{refEme,refCGprl} with experiments in that it
fails to yield a tetracritical point in the $H$-$T$ plane for ${\bf
H}\parallel{\bf\hat c}$ \cite{refgc}. ( Whether there is or is not a
{\it true} tetracritical point in the experimentally observed phase
diagram is irrelevant.  At the very least one has two phase boundaries
$H_{c2}(T)$ and $H_{ci}(T)$ that approach within 10-15~mK of one
another, and the outermost line $H_{c2}(T)$ has a sharp change in the
sign of the curvature near this apparent tetracritical point. The
theory can not explain this `near collision' either.) This led us to
propose the $AB$ model which does not have this flaw. Our proposal in
turn led Sauls to propose a refinement of the $E$ model\cite{refSauls},
which has three ingredients: (i) a specific $E_{2u}$ order parameter
(see below), (ii) a nearly cylindrically symmetric Fermi surface, and
(iii) coupling of the gradient terms in the GL free energy to the SBF.
A cylindrical Fermi surface causes the dangerous gradient coupling
terms in the free energy \cite{refgc} to vanish within weak coupling
BCS theory for an $E_{2}$ order parameter, and in conjunction with
ingredient (iii) restores the tetracritical point for ${\bf
H}\parallel{\bf\hat c}$.

In this paper we shall use GL theory to calculate $H_{c3}$ for UPt$_3$
within the original $E$ model and the $AB$ model. The GL free energy
for the refined $E$ model is mathematically isomorphic to, and can be
regarded as a special case of, the $AB$ model free energy\cite{refd}.
Since the key issue that determines $H_{c3}$ is the boundary condition
on the order parameter, however, the refined $E_{2u}$ model of Sauls
will have the same $H_{c3}(T)$ curves as the unrefined $E_{2u}$ model
for ${\bf H}\perp {\bf\hat c}$. Indeed, $H_{c3}$ curves may be similar
for these two models even when  ${\bf H}\parallel {\bf\hat c}$, except
that one must beware of kinks in the refined model. We will limit
ourselves to cases where the field ${\bf H}$ and the surface normal
${\bf\hat n}$ lie in high crystal symmetry directions or planes.
Further, we only consider ideal, or specularly reflecting surfaces.

The motivation for this study is that we expect qualitative differences
in the behavior of $H_{c3}$ between the various models for various
geometries. Firstly, the boundary conditions are different amongst the
$E$ models, leading to differences in $H_{c3}$. Secondly, in the $AB$
model the eigenvalue equations for $H_{c3}$ decouple into separate
equations for two components, and whenever the surface supports both
components, the $H_{c3}$ curve is expected to mirror $H_{c2}$ and show
a kink. In the $E$ model, on the other hand, even for ${\bf
H}\perp{\bf\hat c}$, the gradient terms can couple the two components.
It is thus possible for the kink in $H_{c3}$ to be smoothed out.  We
shall see that whether this happens or not is a question of dynamics,
not symmetry. In fact, there turns out to be no smoothing with the
gradient coupling values we use. A short quantitative estimate of this
effect seems to be hard to get, however, so we present the full
analysis which follows.  Since an experimental measurement of $H_{c3}$
has now been reported \cite{refKel} for some of the geometries that we
study, we can use our results to restrict the acceptable order
parameters for UPt$_3$.  We also hope that our work will spur a more
detailed experimental study of $H_{c3}$ in other geometries as well, as
this will sharpen our understanding of the order parameter even
further.

We note here that as this paper was being written, we learned of a
recent paper by Samokhin\cite{refSam}, also discussing $H_{c3}$ in
UPt$_3$ within $E$ models. Samokhin gives a microscopic foundation to
the boundary conditions on the order parameter\cite{refBar}, while we
take a purely phenomenological approach. Further, he focuses on the
case where ${\bf H}\parallel {\bf\hat c}$, and the surface normal is
arbitrarily oriented in the $a$-$b$ plane. We have largely avoided
detailed study of the ${\bf H}\parallel{\bf\hat c}$ geometry because of
the problems it presents in comparing to the observed bulk upper
critical field. Thus Samokhin's work is nicely complementary to ours.

The rest of the paper is organized as follows. In Sec. II, we
recapitulate the bare essentials of the theory of surface
superconductivity for fully symmetric superconductors with a complex
scalar order parameter, paying special attention to the boundary
conditions. We then extend these ideas to the $AB$ and $E$ models. The
case of the $E$ model is rather rich. Depending on the exact order
parameter and field and surface orientation, the surface may or may not
suppress superconductivity. We tabulate these cases and proceed to
study them in Secs. III and IV. Certain technical aspects of the
calculations are given in the Appendix. In Sec. V, we compare our
results with the experiments of Keller {\it et. al.} \cite{refKel}, and
see which order parameters are compatible. Our results are summarized
in Table III, and the reader who is not interested in the details of
the analysis should skip to that directly. We conclude with suggestions
for future work.

\section{Basic Theory of surface superconductivity}
\subsection{Fully symmetric superconductors}
We first recapitulate the theory for a superconductor with a complex
scalar order parameter $\psi$, occupying the half space $z\ge
0$\cite{refdeg,refSaiB,refSaint}. $H_{c3}$ is found by solving the
linearized GL equation
\begin{equation}
{1 \over 2m}\left(-i\hbar {\bf \nabla} - {2e \over c}{\bf A}\right)^2
   \psi = -\alpha(T)\psi
\label{eqa}
\end{equation}
subject to the boundary condition of no current flow across the
surface: ${\bf j} \cdot {\bf\hat z} = 0$. The notation in
Eq.~(\ref{eqa}) is standard.

If ${\bf H} \parallel {\bf\hat z}$, then then there is no surface
superconductivity, and technically $H_{c3}=H_{c2}$. The problem is more
interesting when ${\bf H} \perp {\bf\hat z}$. For an interface with a
vacuum or insulator, the boundary condition ${\bf j} \cdot {\bf\hat z}
= 0$ becomes
\begin{equation}
D_z\psi|_{z = 0} = 0. \label{eqb}
\end{equation}
where ${\bf D}$ is the gauge covariant derivative.
Taking ${\bf A} = -Hz {\bf\hat y}$, and writing $\psi$ as a
plane wave in the $x$-$y$ plane with wavevector $(k_x,k_y)$ times a
function $f(z)$, Eqs.~(\ref{eqa}) and (\ref{eqb}) can be rewritten as
\begin{equation}
\left[-{d^2 \over dz^2}+\left({2\pi H \over \Phi_0}\right)^2
   (z-z_0)^2 + k_x^2\right]f = {1\over \xi^2(T)} f, \label{eqc}
\end{equation}
with $(df/dz)_{z=0} = 0$. Here, $z_0 = -k_y\Phi_0/2\pi H$, $\Phi_0$ is
the flux quantum, and $\xi(T) = - \hbar^2/2m\alpha(T)$. The highest
field eigenvalue is obtained with $k_x = 0$, and
\begin{equation}
z_0 = g_0 \xi(T), \quad \quad H_{c3} = {1\over g_0} H_{c2}
\end{equation}
with $g_0 = 0.59010$. $(1/g_0 = 1.6946.)$ Recall that $H_{c2} = \hbar
c/2e\xi^2(T).$

Kittel's variational solution to the above problem is also worth
reproducing. Taking
\begin{equation}
f = (2/\pi\sigma^2)^{1/4} \exp(-z^2/4\sigma^2) \label{eqba}
\end{equation}
and minimizing the total energy with respect to $z_0$ and $\sigma$
gives
\begin{eqnarray}
&& \sigma^2 = {1\over 2g_K}\xi^2(T), \nonumber\\
&& z_0 = (2/\pi)^{1/2}\sigma = 0.727\xi(T), \\
&& H_{c3} = g_K^{-1} H_{c2}, \nonumber
\end{eqnarray}
with $g_K = (1 - 2/\pi)^{1/2} = 0.603$. ($1/g_K = 1.66.$) Note that
this variational solution has the property of zero integrated current
in the $y$ direction:
\begin{equation}
\int_0^\infty j_ydz \propto \int_0^\infty(z-z_0)f^2(z)dz = 0.
\label{eqint}
\end{equation}

Finally, we note that if Eq.~(\ref{eqb}) is replaced by the boundary
condition $\psi(0) = 0$, the highest $H$ eigenvalue is obtained by
letting $z_0 \rightarrow \infty$ in Eq.~(\ref{eqc}), and $H_{c3}
=H_{c2}$.

\subsection{Reduced symmetry superconductors}
For both $E$ and $AB$ models, the order parameter can be written as a
two-component complex vector
\begin{equation}
\eta = \left( \matrix{\eta_1 \cr \eta_2}\right). \label{eqbb}
\end{equation}
This transforms as a single irreducible ($E$) representation of
$D'_{6h}$ in the $E$ model and as a reducible representation in the
$AB$ case. The quadratic part of the GL free energy density (which is
all that matters for $H_{c3}$) can be written compactly as
\begin{equation}
f^{(2)}_{\rm GL} = \sum_{r=1,2} \alpha_r(T)|\eta_r|^2 +
   \sum_{i,j,r,s} \kappa^{rs}_{ij}(D_i\eta_r)^*(D_j\eta_s).
\label{eqAB}
\end{equation}
where $i,j \in \{x,y,z\}$, and the $\kappa$'s are gradient coefficients
suitably constrained by symmetry.

It follows from Eq.~(\ref{eqAB}) that the current is given by
\begin{equation}
j_i = {4e\over \hbar} {\rm Im} \sum_{j,r,s}\kappa^{rs}_{ij}
   \eta^*_rD_j\eta_s.
\end{equation}
At first sight the condition ${\bf j} \cdot {\bf\hat n} = 0$ cannot
be written as a linear equation in $\eta$.  Just as in the case of
ordinary superconductors, however, a linear equation can be
obtained\cite{refSam}, as was first done for $^3$He-A\cite{refamb}.  In
contrast to ordinary superconductors, the roughness of the interface is
now very important. A rough surface is in general pair breaking. Since
we wish to study the maximum possible enhancement of $H_{c3}$, we will
only consider ideal or specularly reflecting surfaces. Further, as
mentioned before, we shall only consider cases where the surface normal
${\bf\hat n}$ and the field ${\bf H}$ lie in high symmetry directions
or planes. Under these conditions, the microscopic analyses
\cite{refSam,refBar,refamb}, which we shall not repeat, show that
the boundary condition on each
component of $\eta$ reduces to \cite{refe}
\begin{equation}
\eta_r = 0 \;\; {\rm or} \;\; ( {\bf\hat n} \cdot {\bf D})\eta_r = 0.
\label{eqe}
\end{equation}
Note that both components need not obey the same condition; one could
have $\eta_1 = 0$ and $( {\bf\hat n} \cdot {\bf D}) \eta_2 = 0$. The
general rule can be written as follows.  Let us write the
momentum-space gap function in the Balian-Werthamer notation,
\begin{eqnarray}
\Delta_0({\bf k}) &=& \sum_{r = 1}^2 \eta_r\psi_r({\bf k}),
  \quad {\rm (even\; parity)}, \\
\vec d({\bf k}) &=& \sum_{r = 1}^2 \eta_r\vec \psi_r({\bf k}),
  \quad {\rm (odd\; parity)},
\end{eqnarray}
where $\psi_r$ or $\vec\psi_r$ are basis functions of the appropriate
representation. Note that in the odd parity case, $\vec d$ and $\vec
\psi_r$ are vectors in pseudo-spin space. The precise boundary
conditions in Eq.~(\ref{eqe}) are then as follows.
Let us denote by ${\sf R}
{\bf k}$ the wave vector obtained from ${\bf k}$ by reflection in the
plane normal to ${\bf\hat n}$. Then, in the even parity case,
at the surface we have,
\begin{eqnarray}
\eta_r &=& 0 \quad{\rm if}\quad \psi_r({\sf R} {\bf k}) =
                             -\psi_r({\bf k}), \label{eqbc1}\\
    ({\bf\hat n \cdot D}) \eta_r &=& 0 \quad{\rm if}\quad
          \psi_r({\sf R} {\bf k}) = +\psi_r({\bf k}). \label{eqbc2}
\end{eqnarray}
The surface acts in effect as a momentum space mirror.
If this mirror plane is a nodal plane for $\psi_r$, then $\psi_r$
must vanish at the surface. If the mirror plane is antinodal, then
the normal derivative must vanish.
In the odd parity case, we replace $\psi_r$ by
$\vec \psi_r$. As noted in Ref.\cite{refYG}, it is unlikely that in
reality $\vec \psi_r$ will be exactly odd or even under reflection
normal to any ${\bf\hat n}$, even those of high symmetry, as the
surface will in general flip pseudospin. Keeping our goal of studying
surface superconductivity under idealized or optimal conditions in
mind, we shall ignore this complication.

In Table I we give a few illustrative cases of boundary conditions
obtained by applying the above rule to candidate order parameters for
UPt$_3$.  Since we will always take ${\bf H} \cdot {\bf\hat n} = 0$,
it is always possible to choose ${\bf A} \perp {\bf\hat n}$, and
the boundary condition (\ref{eqbc2}) reduces to
$\nabla_n\eta_r \equiv \eta'_r = 0$. Several points about this table
should be noted: (i) The directions $x$, $y$, $z$ are fixed along the
crystal symmetry directions with ${\bf\hat x} = {\bf\hat a}$, ${\bf\hat
y}={\bf\hat a}^*$, and ${\bf\hat z} = {\bf\hat c}$.  (ii) For odd
parity, unit vectors in pseudospin space are denoted by a subscript
`s'. (iii) For each representation we have generally given the simplest
possible basis function. As noted in Ref.\cite{refYG} this does not
yield the most general gap function allowed by symmetry. As a
counter-example, we show two order parameters for $E_{2u}$\cite{reff},
distinguished by superscripts. The second of these is in fact the one
advanced by Sauls\cite{refSauls} and by Norman\cite{refg}. (iv) In each
case, we list the pair of quantities that must vanish at the
interface.  Thus for the $E_{2g}$ example, with ${\bf\hat n} =
{\bf\hat x}$, we must have $\eta'_1 = \eta_2 = 0$. (v) In some cases, the
boundary conditions are mixed, i.e. of the type discussed in
Refs.\cite{refSam,refe}.

The reader can write down the correct boundary conditions for any other
order parameter using our rules. In subsequent sections, we will solve
the $H_{c3}$ problem for both $AB$ and $E$-type models. In the case of
the $AB$ models, $H_{c3}$ is independently found for the $A$ and $B$
component. By a simple rescaling of the coordinates, the eigenvalue
equation can be reduced to that of the isotropic case. The boundary
condition is either $\eta_r = 0$ or $\eta'_r = 0$, yielding
$H_{c3}/H_{c2}$ equal to 1 or $g_0^{-1}$, respectively. Having found
$H_{c3}(T)$ in this manner for each component separately, the
thermodynamic $H_{c3}(T)$ is taken to be the larger of the two for each
value of $T$. A similar procedure will turn out to work in many cases
for the $E$-models, although it is not obvious a priori that the
equations for $\eta_1$ and $\eta_2$ will decouple.

\section{$H_{\lowercase{c}3}$ for $AB$ models}
In this section we will calculate $H_{c3}$ for the $AB$ model order
parameters listed in Table I. Writing $(\eta_a, \eta_b)$ instead of
$(\eta_1, \eta_2)$, the free energy (to quadratic order in
$\eta$) for all $AB$ models is given by
\begin{equation}
f_{\rm GL}^{(2)} = \sum_{r=a,b}\left[ \alpha_r |\eta_r|^2 +
     \kappa_r|{\bf D}_\perp \eta_r|^2 +
     \kappa'_r|{\bf D}_z\eta_r|^2\right]. \label{eqfa}
\end{equation}
Here $\alpha_r = \alpha_0(T-T_r)$, and all $\kappa$'s are positive. We
assume that  $\eta_a$ and $\eta_b$ belong to the $A$ and $B$
representations, respectively.

\subsection{${\bf\hat n} = {\bf\hat z}$}

Since all directions of ${\bf H}\perp{\bf\hat n}$ are equivalent, let
us take ${\bf H} = H {\bf\hat x}$, ${\bf A} = - Hz {\bf\hat y}$.
Minimization of $f_{\rm GL}^{(2)}$ leads to the GL equations, which
decouple for $\eta_a$ and $\eta_b$:
\begin{equation}
-\kappa_r\left[{\partial^2 \over \partial x^2} +
   \left({\partial \over \partial y} +
    {2ie\over \hbar c} Hz\right)^2\right]\eta_r -
   \kappa'_r {\partial^2 \over \partial z^2} \eta_r =
     -\alpha_r\eta_r. \label{eqg}
\end{equation}
The boundary conditions are $\partial_z\eta_a = \eta_b = 0$ at $z=0$.

Equation~(\ref{eqg}) can clearly be made isotropic by a rescaling of
coordinates. We then get
\begin{eqnarray}
H_{c3}^a(T) &=& 1.695 H_{c2}^a(T), \\
H_{c3}^b(T) &=& H_{c2}^b(T),
\end{eqnarray}
where the superscript labels the order parameter component that is
nucleating.

Taking the $H_{c2}(T)$ curves to have a kink as seen experimentally, we
obtain a very simple picture for $H_{c3}$. For the even parity case, if
$T_a > T_b$, there is only one branch to $H_{c3}(T)$, and it onsets at
$T_a$, i.e., the upper transition in zero field. If $T_b > T_a$, there
is again only one branch, but it onsets below the upper transition
temperature, and it extrapolates to $T_b$ for $H=0$. These
possibilities are shown schematically in Figs. 1 and 2. For the odd
parity cases, the roles of $\eta_a$ and $\eta_b$ are reversed, i.e.,
Fig. 1 applies if $T_b > T_a$, and Fig. 2 applies if $T_a > T_b$.

\subsection{${\bf\hat n}={\bf\hat x}$ or ${\bf\hat n} = {\bf\hat y}$}

Let us first consider  the $A_{1g}\oplus B_{2g}$ order parameter. If we
take ${\bf\hat n} = {\bf\hat x}, {\bf H} = H {\bf\hat y}$, the
analysis is identical to that just given, and we recover Figs. 1 and 2
for $T_a > T_b$ and $T_b > T_a$ respectively. If ${\bf\hat n} =
{\bf\hat y}$, and $H = H {\bf\hat x}$, say, then taking $\vec A = Hy
{\bf\hat z}$, the GL equation becomes
\begin{equation}
-\kappa_r\left({\partial^2 \over\partial x^2} +
   {\partial^2\over\partial y^2}\right)\eta_r -
    \kappa'_r\left({\partial\over \partial z} -
    {2ie\over\hbar c} Hy\right)^2\eta_r = -\alpha_r \eta_r,
\end{equation}
which can be isotropized as before, but the boundary conditions now are
$\partial_z \eta_a = \partial_z \eta_b = 0$ at $z=0$, so $H_{c3}$ is
given by
\begin{equation}
H_{c3}^r(T) = 1.695 H_{c2}^r(T), \;\; r = a,b.
\end{equation}
The resulting picture is shown in Fig.~3. Note that within the GL
approximation, the $H_{c3}$ and $H_{c2}$ kinks occur at the same
temperature, as can be proved by simple geometry. Further, in this
case, we expect to see inner transition lines analogous to those for
$H_{c2}$, corresponding to the onset of surface superconductivity in
the other component.

We get the same answers as above for the $A_{1u} \oplus B_{1u}$ case,
as the boundary conditions are the same. For the $A_{1u} \oplus B_{2u}$
order parameter, the results for the ${\bf\hat n} ={\bf\hat x}$ and the
${\bf\hat n} ={\bf\hat y}$ cases are interchanged.

\section{$H_{\lowercase{c}3}$ for $E$ models}
We turn at last to our main problem, namely that of $H_{c3}$ for the
$E$ models. For the reasons mentioned in Sec.~I, we will not consider
explicitly the refinements made by Sauls\cite{refSauls}.

The free energy for all $E$ models is more conveniently represented by
regarding $(\eta_1,\eta_2)$ as the components of a vector
in the $x$, $y$ plane. This is self-evident for the $E_1$ cases, but as
shown by Tokuyasu\cite{reftaku}, the number and form of the invariants
is the same for the $E_2$ cases as well. Writing $(\eta_1,\eta_2)$ as
$(\eta_x,\eta_y)$, we can write the quadratic part of $f_{\rm GL}$ as
\begin{equation}
f_{\rm GL}^{(2)} = \alpha_+|\eta_y|^2 +
  \alpha_-|\eta_x|^2 + \kappa_1 D^*_i\eta^*_jD_i\eta_j +
  \kappa_2D^*_i\eta^*_iD_j\eta_j+\kappa_3D^*_i\eta^*_jD_j\eta_i +
  \kappa_4D^*_z\eta^*_iD_z\eta_i, \label{eqh}
\end{equation}
where, $i,j \in \{x,y\}$, and we sum over repeated indices. Also, we
have $\alpha_\pm = \alpha_0(T-T_\pm)$, with $T_\pm = T_{c0} \pm
\epsilon/\alpha_0$.

We write the GL equations implied by Eq.~(\ref{eqh}) using the notation
of Ref.~\cite{refh}. We scale all lengths by $l = (\hbar c/2eH)^{1/2}$,
$\alpha_0T$ and $\epsilon$ by $\kappa_b/l^2$ with $\kappa_b = \kappa_1
+ (\kappa_{23}/2)$, and $\kappa_{23} = \kappa_2 + \kappa_{3}$. We also
define $p_i = -iD_i$, $u = \kappa_{23}/2\kappa_b$, $v = (\kappa_2 -
\kappa_3)/2\kappa_b$, $w = \kappa_4/\kappa_b$. Then, the GL equation is
\begin{equation}
H_{\rm GL} \left(\matrix{\eta_x \cr \eta_y}\right) =
  E\left(\matrix{\eta_x\cr\eta_y}\right),
\end{equation}
with
\begin{equation}
H_{\rm GL} = \left[\matrix{(1+u)p_x^2 + (1-u)p_y^2+wp_z^2 +
     \tilde\epsilon & u\{p_x,p_y\} + ivH_z/H \cr
  u\{p_x,p_y\} + ivH_z/H & (1-u)p_x^2 + (1+u)p_y^2+wp_z^2 -
     \tilde\epsilon}\right]. \label{eqi}
\end{equation}
Here, $\{p_x,p_y\}$ is an anticommutator,
$\tilde\epsilon = \epsilon l^2/\kappa_b$, and
$E = \alpha_0 (T - T_{c0}) l^2/\kappa_b$. Note in particular that
$\tilde\epsilon \propto 1/H$.

To avoid misunderstanding, we emphasize that in this paper we will
always take the axes $x$, $y$, and $z$ as fixed along the crystal axes
$a$, $a^*(\perp a$), and $c$. This convention differs from that adopted
by many other workers who are guided by the near isotropy of $H_{c2}$
for basal plane fields \cite{refKel,refSauls2,refAg,refKel2}.  This
effect is generally explained by assuming that the staggered
magnetization ${\bf M}^\dagger$ rotates so as to be orthogonal to ${\bf
H}$. It is then convenient to rotate the axes, and the pair $(\eta_x,
\eta_y)$ so that $\bf H$ is always along ${\bf\hat x}$ (or ${\bf\hat
y}$), and ${\bf M}^\dagger$ is implicitly along ${\bf\hat y}$ (or
${\bf\hat x}$).  We shall not do this. Rather, we shall rotate $\bf H$
and ${\bf M}^\dagger$, and use a fixed $k$-space form for $\eta_x$ and
$\eta_y$. The coupling between ${\bf M}^\dagger$ and $\eta$ then
changes as ${\bf M}^\dagger$ rotates, and for ${\bf M}^\dagger$ along
the ${\bf\hat x}$ or ${\bf\hat y}$ axes, in particular, it takes the
form
\begin{equation}
F_{\rm SBF} = \cases{-\zeta|{\bf M}^\dagger|^2(|\eta_x|^2 -
    |\eta_y|^2), \quad {\bf M}^\dagger \parallel {\bf\hat x}, \cr
   -\zeta|{\bf M}^\dagger|^2(|\eta_y|^2 - |\eta_x|^2), \quad
    {\bf M}^\dagger \parallel {\bf\hat y}, \cr}
\end{equation}
where $\zeta$ is a coupling constant. Thus, strictly speaking, with our
convention, Eq.~(\ref{eqi}) is correct as written only if $\bf H$ lies
in the $x$-$z$ or $y$-$z$ planes, and the sign of $\tilde\epsilon$
depends on that of $\zeta$. The latter depends on presently unknown
microscopic physics. It can be seen, however, that Eq.~(\ref{eqi}) has
a symmetry with respect to a change in sign of $u$. By taking ${\rm
sgn}(u) = {\rm sgn}(\zeta)$, we can explain the observed kink in
$H_{c2}$, whatever  ${\rm sgn}(\zeta)$ may be.

To calculate $H_{c3}$, we must add boundary conditions. Before
discussing these, however, let us comment further on the parameters
$u$, $v$, and $w$. Firstly, local stability of the uniform solution
requires that\cite{refEmb}
\begin{equation}
w>0,\;\; 1-v>0, \;\; 1+v > 2|u|.
\end{equation}
It follows from these that $|u| < 1$. Secondly, the ${\bf
H}\perp{\bf\hat c}$ case involves $w$ and $u$, but only $|u|$ can be
found from experimental $H_{c2}$ slopes. (The ${\bf H}\parallel{\bf\hat
c}$ case involves only $v$, but comparison with experiment is
problematic as discussed in Section I.) The values of $u$ so determined
vary quite a bit. We shall take $|u|=0.46$ in our numerical work in
accord with the data of Ref.~\cite{refPDa}.

\subsection{${\bf\hat n} = {\bf\hat z}$, ${\bf H}\perp {\bf\hat z}$}

Because of basal plane isotropy, we expect the same $H_{c3}$ curve for
all field orientations in this case. Let us take ${\bf H}\parallel
{\bf\hat x}$, $\epsilon > 0$, and let us work in the gauge ${\bf A} =
-Hz {\bf\hat y}$. Then in Eq.~(\ref{eqi}), $p_z = -i\partial_z$, $p_x =
-i\partial_x$, $p_y = -i\partial_y + z$. We can clearly take $\eta$ as
a plane wave in the $x$-$y$ plane and replace $p_x$ by $k$ and $p_y$ by
$z_0-z$, where $k$ and $-z_0$ are wavevector components, Then,
$\{p_x,p_y\} \rightarrow 2k(z-z_0)$, and we get a one-dimensional
Hamiltonian,
\begin{equation}
H_{\rm GL} = \left[\matrix{-w\partial^2_z+(1-u)(z-z_0)^2+(1+u)k^2 +
    \tilde\epsilon & 2ku(z-z_0) \cr 2ku(z-z_0) &
   -w\partial^2_z+(1+u)(z-z_0)^2+(1-u)k^2 -\tilde\epsilon}\right].
\label{eqj}
\end{equation}

For the $E_{2g}$, $E_{1u}$, and $E_{2u}^{(1)}$ cases listed in Table I,
the boundary conditions are $\partial_z\eta_x = \partial_z\eta_y = 0$
at $z=0$.  If $k=0$ for all $T$, we get two decoupled $H_{c3}$ problems
and a kink in $H_{c3}$ as in Fig.~3. The only nontrivial features arise
from the possibility of having $k \not= 0$. Such $k$-mixing is found to
affect the $H_{c2}$ curves for $u\ge 0.61$ for basal plane fields
\cite{refh}. Whether the lowest energy eigenvalue for the $H_{c3}$
problem ever corresponds to $k \not= 0$ is a dynamical question, to
which we have not been able to find a two-line answer.  It is obvious
that having $k\not= 0$ raises the energy from the diagonal terms in
Eq.~(\ref{eqj}). The compensation of this rise from the off-diagonal
terms is likely to be largest when the $x$ and $y$ solutions are nearly
degenerate, i.e. for fields near the kink found with $k=0$.
If the energy were lowered by having $k\not= 0$, the crossing of the $x$
and $y$ solutions would turn into an anticrossing, and the kink would be
smoothed out. Our goal is to explore this possibility.

It is clear from Eq.~(\ref{eqj}), that a $k\not= 0$ solution is more
favored for large values of $u$. Let us mention two obvious analytic
approaches to this problem. Firstly, a lower bound on the minimum $u$
required can be found by constructing a Schwarz inequality for the
matrix element of the off-diagonal term in Eq.~(\ref{eqj}), but this
bound is too small to be useful. Secondly, one might try and construct
variational solutions analogous to Eq.~(\ref{eqba}), with different
widths $\sigma_x$ and $\sigma_y$ for $\eta_x$ and $\eta_y$.  The energy
is then a function of five variational parameters, $\sigma_x$,
$\sigma_y$, $z_0$, $k$, and a mixing angle, and the minimization is
complicated enough that one might as well attempt a nonvariational
numerical solution. This approach does reveal one fact, however. If $u$
is small, $\sigma_x$ and $\sigma_y$ are likely to be similar, and
Eq.~(\ref{eqint}) shows that the matrix element of $(z-z_0)$ is likely
to be small. Thus the effect of the off-diagonal terms in
Eq.~(\ref{eqj}) is probably of $O(u^2)$.

We have therefore solved the eigenvalue problem for $H_{c3}$
numerically. We do this by working in a basis of harmonic oscillator
eigenfunctions with unequal length scales $(w/(1\mp u))^{1/4}$ for
$\eta_x$ and $\eta_y$.  Only even parity (about $z=0$) functions are
used in accord with the boundary conditions. The Hamiltonian
(\ref{eqj}) is almost diagonal in this basis. The evaluation of the
off-diagonal terms is described in the Appendix. We find the lowest
eigenvalue $E_{\rm min}$ and associated eigenfunction $|\eta_{\rm
min}\rangle$ of $H_{\rm GL}$ numerically for given $k$ and $z_0$, and
minimize with respect to $k$ and $z_0$ via the conjugate gradient
method. The required gradients are efficiently found using the identity
\begin{equation}
{\partial\over\partial k} E_{\rm min} = \biggl\langle\eta_{\rm min}
\biggl |{\partial H_{\rm GL}\over\partial k}\biggr|
\eta_{\rm min}\biggr\rangle,
\end{equation}
and a similar one for the $z_0$ derivative. These calculations are done
for $\tilde\epsilon$ ranging from 0.1 to 10. This range is large enough
to encompass the kink region comfortably, as all terms in
Eq.~(\ref{eqj}) are of order unity.

We find that for $u = 0.46$, $k$ is numerically equal to zero for all
$\tilde \epsilon$ values considered. In fact, $k$-mixing occurs only
for $u\ge 0.59$, but the resulting anticrossing is very narrow, and
substantial curvature in $H_{c3}(T)$ is only visibly evident for
$u\ge 0.8$. Thus, for experimentally relevant values of $u$,
the picture for $H_{c3}$ is
the same as in Fig. 3, i.e., the $H_{c3}/H_{c2}$ ratio is always ideal
and there is a kink in $H_{c3}$\cite{refj}. Once again, we expect to
see inner $H_{c3}$ lines which we have not explicitly calculated.

The $H_{c3}$ problem is trivial for the $E_{1g}$ and the $E_{2u}^{(2)}$
cases listed in Table I. The boundary conditions now are $\eta_x =
\eta_y = 0$ at $z=0$, and the surface cannot support superconductivity,
i.e., $H_{c3}=H_{c2}$.

\subsection{${\bf\hat n} \perp {\bf\hat z}$, ${\bf H}\parallel
{\bf\hat z}\times{\bf\hat n}$}

In this geometry, as can be seen from Table I, except for the
$E_{2u}^{(1)}$ order parameter, the boundary conditions call for only
one of the $\eta_i$ to vanish at the surface. The other component is
not suppressed by the surface. The form of $H_{c3}(T)$ is different
depending on whether the surface suppressed component has the higher or
lower transition temperature. This in turn depends on the sign of the
coupling of the symmetry breaking field (SBF) to the order parameter.

To understand this point, let us first consider the $E_{1g}$ and
$E_{1u}$ cases from Table I. The two high-symmetry geometries are shown
in Fig. 4. For cases (a) and (b), we have ${\bf M}^\dagger
\parallel{\bf\hat x}$ or ${\bf M}^\dagger\parallel{\bf\hat y}$,
and $\epsilon =
-\zeta|{\bf M}^\dagger|^2$ or $\epsilon = +\zeta|{\bf M}^\dagger|^2$,
respectively. Suppose $\zeta > 0$. Then, for the case of Fig. 4(a) we
have $\eta'_y = 0$ and $\epsilon < 0$, i.e., $\eta_y$ has the lower
$T_{c}$. For the case of Fig. 4(b) we have $\eta'_x = 0$ and $\epsilon
> 0$, i.e., $\eta_x$ has the lower $T_{c}$. In both geometries, the
surface and the SBF act in opposition, i.e., the SBF is oriented so
that the surface supported order parameter component has the lower
$T_{c}$. In contrast, if $\zeta < 0$, then the surface and the SBF act
in concert, and the surface supported component has higher $T_{c}$.

The sign of $\zeta$ is unknown and fixed by microscopic physics. We can
conclude, though, that whether the surface and SBF act in concert or
opposition, for $E_1$ order parameters, the same behavior will be seen
for the two geometries in Fig. 4. This point can be understood as
follows. An $E_1$ order parameter behaves as a vector in the $x$-$y$
plane, and so if we rotate the surface normal ${\bf\hat n}$, keeping
${\bf H}\parallel {\bf\hat z}\times{\bf\hat n}$, the boundary
conditions on $\eta$ rotate in the same way, and we get an isotropic
$H_{c3}$.

For the $E_{2g}$ and $E_{2u}^{(2)}$ order parameters, we can see from
Table I that the same component, $\eta_x$, is supported by the surface
for both ${\bf\hat n} = {\bf\hat x}$ and  ${\bf\hat n} = {\bf\hat y}$.
For any ${\rm sgn}(\zeta)$, this component must have the lower $T_c$
for one orientation, and the higher $T_c$ for the other, leading to
different answers for $H_{c3}$. In terms of symmetry, this can be
understood by saying that as ${\bf\hat n}$ is rotated in the basal
plane, an $E_2$ order parameter rotates at twice the rate of a vector.
Denoting the angle between ${\bf\hat n}$ and ${\bf\hat x}$ by $\phi$,
the boundary conditions change with $\phi$, reversing every
$30^{\circ}$. Thus the SBF and the surface will act in concert for one
geometry in Fig. 4, and opposition for the other.

We study the cases of surface and SBF in concert and in opposition
separately below. It is convenient to fix $\tilde\epsilon > 0$, so that
$\eta_y$ always has the higher $T_c$, and to rotate ${\bf\hat n}$ and
$\bf H$ as needed.

\subsubsection{Surface and SBF in concert}

We take the geometry of Fig. 4(a), and $\eta'_y = \eta_x = 0$ at $x =
0$. We also take ${\bf A} = - Hx {\bf\hat z}$, allowing us to put $p_z
= (-i\partial _z + x)$, $p_x = -i\partial_x$, $p_y = -i\partial_y$ in
Eq.~(\ref{eqi}). Further taking $\eta$ as a plane wave in the $yz$
plane, we can replace $p_y$ by $k$, $p_z$ by $(x-x_0)$, and
$\{p_x,p_y\}$ by $-2ik\partial_x$.  (Obviously, $k$ and $x_0$ are
wavevector components.) The resulting one dimensional GL Hamiltonian is
\begin{equation}
H_{\rm GL} = \left[\matrix{-(1+u)\partial_x^2 + (1-u)k^2+w(x-x_0)^2 +
    \tilde\epsilon & -2iuk\partial_x \cr
    -2iuk\partial_x & -(1-u)\partial_x^2 + (1+u)k^2+w(x-x_0)^2 -
    \tilde\epsilon}\right].
\end{equation}

If $k=0$, we get two decoupled problems, and because of the boundary
conditions $H_{c3}^y = 1.69 H_{c2}^y$, $H_{c3}^x =  H_{c2}^x$. This
leads to the phase diagram of Fig.~1.

Thus, as in subsection A, the only real question is whether $k\ne 0$
gives a lower energy. We have diagonalized $H_{\rm GL}$ numerically
using a harmonic oscillator wavefunction basis with length scales
$[(1\pm u)/w]^{1/4}$, and odd and even parities about $x=0$, for
$\eta_x$ and $\eta_y$ respectively. The calculation is done as in
subsection A, and we again find that with $u = -0.46$, $k=0$ for all
values of $\tilde \epsilon$ \cite{refkmix}. The $H_{c3}$ picture is
that of Fig.~1.

\subsubsection{Surface and SBF in opposition}

We take the geometry of Fig. 4(b), and $\eta_y=\eta'_x = 0$ at $y=0$.
Further, with ${\bf A} = H y{\bf\hat z}$, and a plane wave dependence
in the $xz$ plane for $\eta$, we get a one-dimensional problem in the
$y$ direction. Numerical work again reveals that $k=0$ for all $\tilde
\epsilon$ \cite{refkmix}, so the $H_{c3}$ picture is as shown in
Fig. 2.

\subsection{${\bf\hat n}\perp {\bf\hat z}$, ${\bf H}\parallel
  {\bf\hat z}$.}

The high-symmetry geometries are shown in Fig.~5. In this case the
field can not orient the SBF. Let us first assume that ${\bf
M}^{\dagger}$ stays fixed parallel to $a^*$, i.e., that the appearance
of superconductivity at the surface does not reorient ${\bf
M}^{\dagger}$. Now, by the argument of the previous subsection, we
expect the same $H_{c3}$ behavior in the two geometries for the
$E_{2g}$ and $E_{2u}^{(2)}$ order parameters, and opposite behavior for
the the $E_1$ order parameters.

The two types of behavior, i.e., surface and SBF in concert and
opposition, can be seen by studying the $E_{1g}$ case with $\zeta<0$.
For Fig.~5(a) and 5(b), we have $\eta_x = \eta'_y = 0$ at $x=0$, and
$\eta'_x = \eta_y = 0$ at $y=0$ respectively. Since ${\bf H}\parallel
{\bf\hat z}$, there is no kink in $H_{c2}$ and none is expected in
$H_{c3}$ either. The ratio $H_{c3}/H_{c2}$ is still of interest and we
have found this numerically. When $\eta_x =0$, [Fig. 5(a)], surface
superconductivity is not supported in the higher $T_c$ component, and
we find, as expected, that $H_{c3}=H_{c2}$.  When $\eta'_x=0$, [Fig.
5(b)], surface superconductivity is supported in the component that
onsets first as $T$ is lowered, and $H_{c3} > H_{c2}$. Our results with
$u=0.46$, and $v=0.1$ are shown in Fig. 6. The ratio $H_{c3}/H_{c2}$
varies from 1.71 for $T\approx T_{c+}$ to 1.77 for $T \approx
T_{c+}-2\Delta T_c$.

The assumption of fixed ${\bf M}^\dagger$ parallel to $a^*$ may not be
physically relevant, however. Since the observed magnitude of ${\bf
M}^{\dagger}$ is very small\cite{refaepp,refGold}, reorientation of
${\bf M}^{\dagger}$ near the surface must be considered. If ${\bf
M}^{\dagger}\parallel {\bf\hat a}$ near the surface, and ${\bf
M}^{\dagger}\parallel {\bf\hat a}^*$ in the bulk, the superconducting
condensation energy could outweigh the magnetic anisotropy and domain
wall energies. In this case, we expect a picture resembling Fig.~6 for
all the $E$ models for any ${\bf\hat n}\perp{\bf\hat z}$, ${\bf
H}\parallel{\bf\hat z}$.

\section{Comparison with experiment, predictions, and conclusions}

As stated before, Keller {\it et al.} have reported\cite{refKel}
surface superconductivity in a whisker of UPt$_3$. The long axis of the
whisker is along the $c$-axis of the crystal, and it has six facets
along the $\{100\}$ crystal planes, i.e., the surface normals are
${\bf\hat n}={\bf\hat x}={\bf\hat a}$, and five others obtained by
successive $60^{\circ}$ rotations about ${\bf\hat z}$. Keller {\it et
al.} study the ac resistivity and susceptibility for various field
orientations, two of which are ${\bf H}\parallel{\bf\hat a}^*$ and
${\bf H}\parallel{\bf\hat c}$, corresponding to the geometries in Figs.
4(a) and 5(a) of our paper. The magnitude of the ratio of the critical
field to the bulk $H_{c2}$ (1.19 and 1.35 for ${\bf H}\parallel{\bf\hat
a}^*$), plus the cusp-shaped deviations in the critical field for
fields with a small (positive or negative) component along ${\bf\hat
n}$ (see Figs. 1 and 2 of their paper), leave little doubt that the
observation is indeed of $H_{c3}$.

Let us begin by studying the case where ${\bf H}\parallel{\bf\hat
a}^*={\bf\hat y}$. Recall that ${\bf\hat n}= {\bf\hat x}$. Keller {\it
et al.} report $H_{c3}/H_{c2}$ = 1.35 for $T= 470$ mK. Given that (i)
$T_{c+} = 508$ mK, and that the bare $T_c$ splitting
$2\epsilon/\alpha_0$, is about 35 mK in the annealed crystals, and (ii)
that $H_{c3}/H_{c2}$ decreases to 1.15 and 1.05 for $T = 450$ and 415
mK, respectively, which are above and below the tetracritical
temperature, $T^* = 430$ mK, we tentatively conclude that the $H_{c3}$
curve is of the type shown in Fig. 1, i.e., there is surface
superconductivity in only one component, and it onsets at $T_{c+}$.
This picture is obtained within the $A_{1g}\oplus B_{2g}$ and
$A_{1u}\oplus B_{1u}$ models if $T_a > T_b$. Within the $A_{1u}\oplus
B_{2u}$ model the predicted picture is that of Fig. 3, whether $T_a >
T_b$ or $T_b > T_a$. This does not seem to fit the observations, which
would appear to exclude the specific $A_{1u}\oplus B_{2u}$ order
parameter listed in Table I. Within the $E$ models, the picture
requires that $\eta_x$ have the higher $T_c$ when ${\bf
H}\parallel{\bf\hat y}$, i.e. ${\bf M}^\dagger\parallel{\bf\hat x}$.
This in turn requires $\zeta < 0$ for the $E_1$ models, and $\zeta > 0$
for the $E_2$ models. These facts are summarized in Table II.

Let us now consider the ${\bf H}\parallel{\bf\hat c}$, ${\bf\hat n} =
{\bf\hat x}$ case. An $H_{c3} > H_{c2}$ is seen at $T=0.932T_c \approx
473$ mK. Since data are not given at any other $T$, the picture of
$H_{c3}(T)$ could resemble any of our Figs. 1, 3, or 6. For the $AB$
models we expect the same qualitative behavior as when ${\bf
H}\parallel{\bf\hat a}^*$, since the direction of {\bf H} in the plane
normal to ${\bf\hat n}$ does not alter the boundary conditions. For any
of the four $E$ models ($E_{1g}$, $E_{2g}$, $E_{1u}$, $E_{2u}^{(2)}$),
on the other hand, a little work shows that if we assume that ${\bf
M}^\dagger$ stays fixed along ${\bf\hat a}^*$, then the sign of $\zeta$
in Table II is such that for the present geometry the surface and SBF
are in opposition. This would imply no surface superconductivity. As
argued in Sec. IV C, however, a reorientation of ${\bf M}^\dagger$ is
quite possible, in which case all four $E$ models would lead to an
$H_{c3}$ curve as in Fig. 6.

We thus see that  the present $H_{c3}$ data can only be used to exclude
(and not very confidently at that) the $A_{1u}\oplus B_{2u}$ order
parameter in Table I. We can however predict the expected behavior for
the other geometries based on the results in Secs. III and IV, given
the restrictions in Table II. These predictions are summarized in Table
III, which is the main result of our paper.

It can be seen that widely different results are obtained for the
geometries ${\bf\hat n} = {\bf\hat c}$ and ${\bf\hat n} =
{\bf\hat a}^*$.
The case ${\bf\hat n}={\bf\hat c}$ distinguishes between
$E_1$ and $E_2$ models and between $E$ and $AB$ models, if one can
confidently detect the presence or absence of a kink in $H_{c3}$. The
case with ${\bf\hat n} = {\bf\hat a}^*$ and ${\bf H}\parallel {\bf\hat
a}$ distinguishes between the odd and even parity $E$ models, and the
$AB$ models. Systematic experimental efforts to study these cases,
would be, to state the obvious, extremely valuable. Even for the one
surface that is presently available, ${\bf\hat n} = {\bf\hat a}$,
studying the variation of $H_{c3}$  with field orientation in the
$a^*$-$c$ plane would be useful and would impel more theoretical
study.

\section*{Acknowledgments}

We are indebted to W. P. Halperin,  Jan Kycia, Jim Sauls, and Sungkit
Yip for many illuminating discussions. This work is supported by the
National Science Foundation through the Northwestern University
Materials Research Center, Grant No. DMR-9120521.

\appendix
\section*{}
We describe here our numerical procedure for finding the matrix of the
GL Hamiltonian (\ref{eqi}), using the case of Sec. IV A for
specificity. We denote the operator valued elements of the matrix
(\ref{eqj}) by $H_{xx}$, $H_{xy}$, $H_{yx}$, and $H_{yy}$, and define
\begin{equation}
H_{ii}^{(0)} = -w\partial^2_z + (1\mp u)z^2,\quad i = x,y.
\end{equation}
We work in the eigenbasis $|n,i\rangle$ of
$H_{ii}^{(0)}$:
\begin{equation}
H_{ii}^{(0)}|n,i\rangle = \left[w(1\mp u)\right]^{1/2}(2n+1)
   |n,i\rangle,\quad i=x,y;\; n=0,1,2,\ldots .
\end{equation}
These functions are normalized on the half-space $z>0$. Depending on
the boundary conditions, only even or odd $n$ may be needed.

The matrix representation of $H_{\rm GL}$ in this basis is clearly very
simple, except for the following elements:
\begin{eqnarray}
P_{nm} &=& \langle n,x|z|m,x\rangle,  \\
Q_{nm} &=& \langle n,x|m,y \rangle, \\
R_{nm} &=& \langle n,x|z|m,y \rangle .
\end{eqnarray}
We do not need to define $\langle n,y|z|m,y\rangle$ separately, as
\begin{equation}
\langle n,y|z|m,y \rangle = [(1+u)/(1-u)]^{1/4}P_{nm}.
\end{equation}

These matrix elements can be found efficiently using the generating
function of the Hermite polynomials. We show how this is done for
$R_{nm}$. Note first, that
\begin{equation}
R_{nm} = {2\over \sqrt{\pi\xi_x\xi_y2^{n+m}n!m!}}
  \int_0^{\infty}z\exp
     \left(-{z^2\over 2\xi_x^2}-{z^2\over 2\xi_y^2}\right)
 H_n\left({z\over \xi_x}\right) H_m\left({z\over \xi_y}\right) dz,
\label{apd}
\end{equation}
where $\xi_{x,y}=[w/(1\mp u))]^{1/4}$, and $H_n$ is a Hermite
polynomial. Denoting the integral in Eq.~(\ref{apd}) by $I_{nm}$, and
using the generating function, we get
\begin{eqnarray}
\sum_{n,m=0}^\infty {I_{nm}s^nt^m\over n!m!} &=&
  \int_0^\infty z  \exp(-s^2+2sz/\xi_x-z^2/2\xi_x^2)
    \exp(-t^2+2tz/\xi_y-z^2/2\xi_y^2) dz \nonumber \\
  &=& -{\sqrt{\pi}\over 2A} e^{(B^2-4AC)/4A}\left[{B\over2\sqrt{A}}
   {\rm erfc}\left({B\over2\sqrt{A}}\right)+
   {\rm erfc}'\left({B\over2\sqrt{A}}\right)\right], \label{ape}
\end{eqnarray}
where $A=(\xi_x^{-2}+\xi_y^{-2})/2$, $B=-2(s/\xi_x+t/\xi_y)$,
$C=s^2+t^2$, and ${\rm erfc'}(x)=d{\rm erfc}/dx$, with
\begin{equation}
{\rm erfc}(x) = 1 - {2\over\sqrt{\pi}}\left(x-{x^3\over3}+
{x^5\over5\cdot2!}- {x^7\over7\cdot3!} + \cdots\right).
\end{equation}

The problem of finding $R_{nm}$ is now reduced to expanding the right
hand side of Eq.~(\ref{ape}) in powers of $s$ and $t$, and comparing
coefficients. This can be done as follows. Consider a polynomial
$A(s,t)$, of maximum order $s^Nt^M$:
\begin{equation}
A(s,t) = \sum_{n=0}^N\sum_{m=0}^M a_{nm}s^nt^m.
\end{equation}
The array $a_{nm}$ defines the polynomial. The array for the sum of two
polynomials $A(s,t)$ and $B(s,t)$ with arrays $a_{nm}$ and $b_{nm}$ is
obviously
\begin{equation}
c_{nm} = a_{nm}+b_{nm},\quad 0\le n\le N,\;0\le m\le M. \label{applus}
\end{equation}
The product $A(s,t)B(s,t)$ has terms up to order $s^{2N}t^{2M}$. If we
are only interested in terms up to order $s^Nt^M$, however, we can
write the corresponding array as
\begin{equation}
c_{nm} =  \sum_{i=0}^n\sum_{j=0}^m a_{ij}b_{n-i,m-j},
  \quad 0\le n\le N,\;0\le m\le M.  \label{approd}
\end{equation}
Note that only array elements $a_{nm}$ and $b_{nm}$ with $n\le N$,
$m\le M$ appear in this formula.

The above procedure can be used to expand the right hand side of
Eq.~(\ref{ape}). Suppose we want $I_{nm}$ for $n,m\le N^*$, where $N^*$
is a number of order 20, say. We write $B$ and $C$ as polynomials in
$s$ and $t$, and represent them by arrays as above. ( Most of the
elements of these arrays are zero at this stage.) We then Taylor expand
$\exp(B^2/4A)$, $\exp(-C)$,  ${\rm erfc}(B/2\sqrt{A})$, and ${\rm
erfc}'(B/2\sqrt{A})$, and repeatedly use the array operations
(\ref{applus}) and (\ref{approd}) with $N=M=N^*$, to evaluate
Eq.~(\ref{ape}). The $I_{nm}$ can be directly read off the resulting
array.

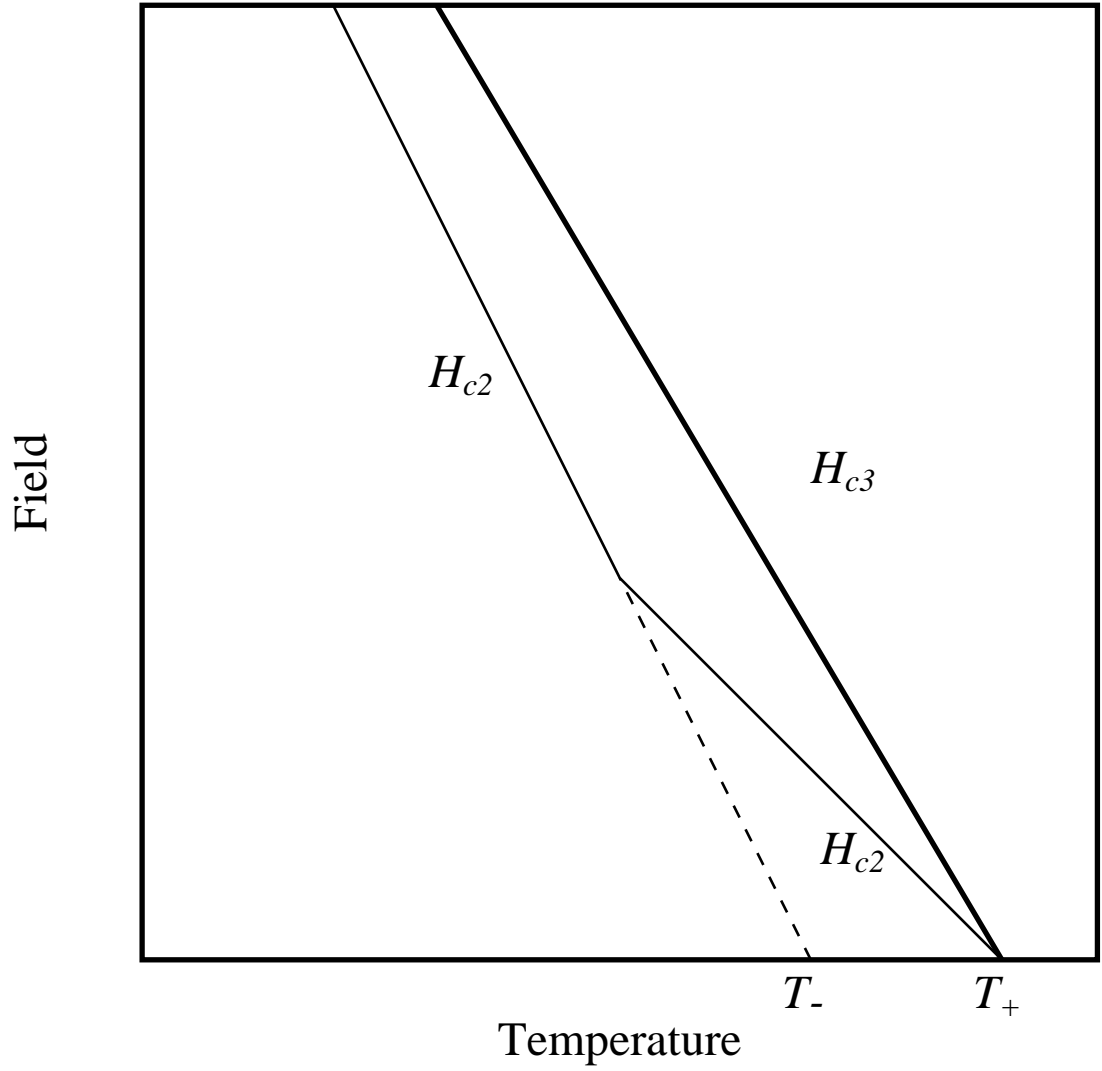
\begin{figure}
\caption{Schematic behavior of $H_{c3}(T)$ when only the higher $T_c$
order-parameter component supports surface superconductivity.}
\end{figure}

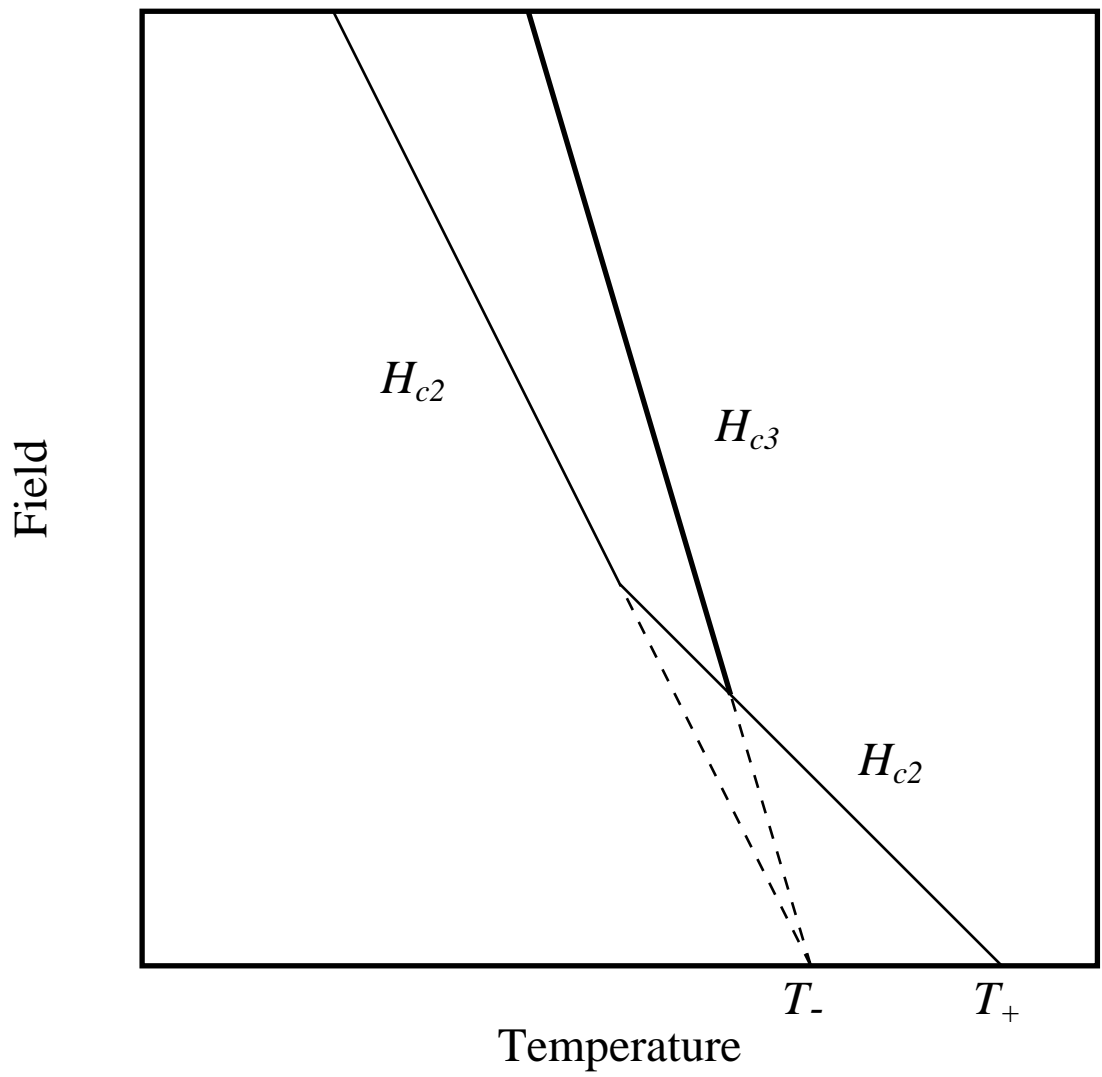
\begin{figure}
\caption{Same as Fig. 1, when only the lower $T_c$ component supports
surface superconductivity.}
\end{figure}

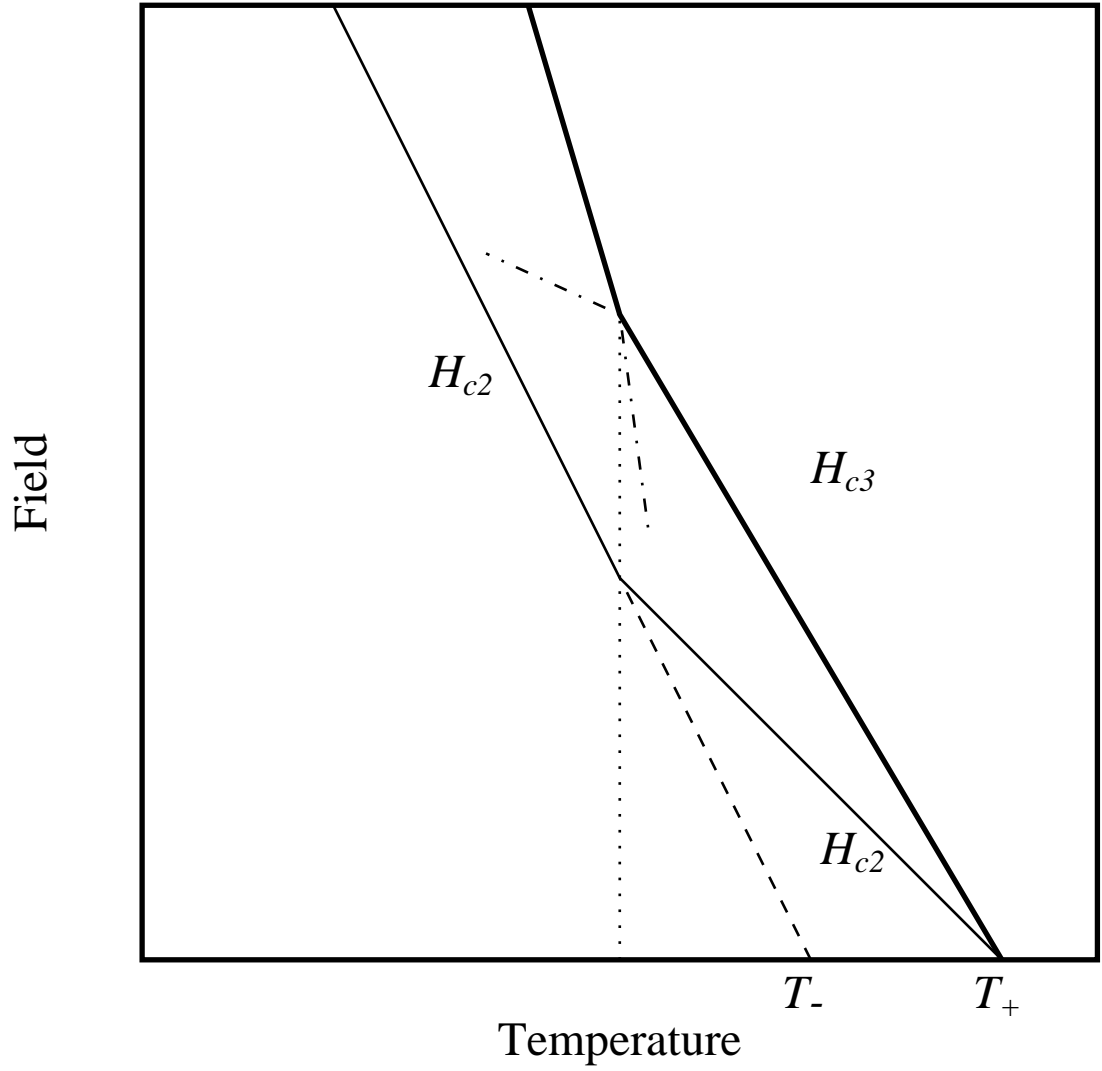
\begin{figure}
\caption{Same as Fig. 1, when both components support surface
superconductivity. In this case we expect inner $H_{c3}$ lines
(dot-dashed) near the kink in $H_{c3}$. We have not explicitly
calculated these lines.}
\end{figure}

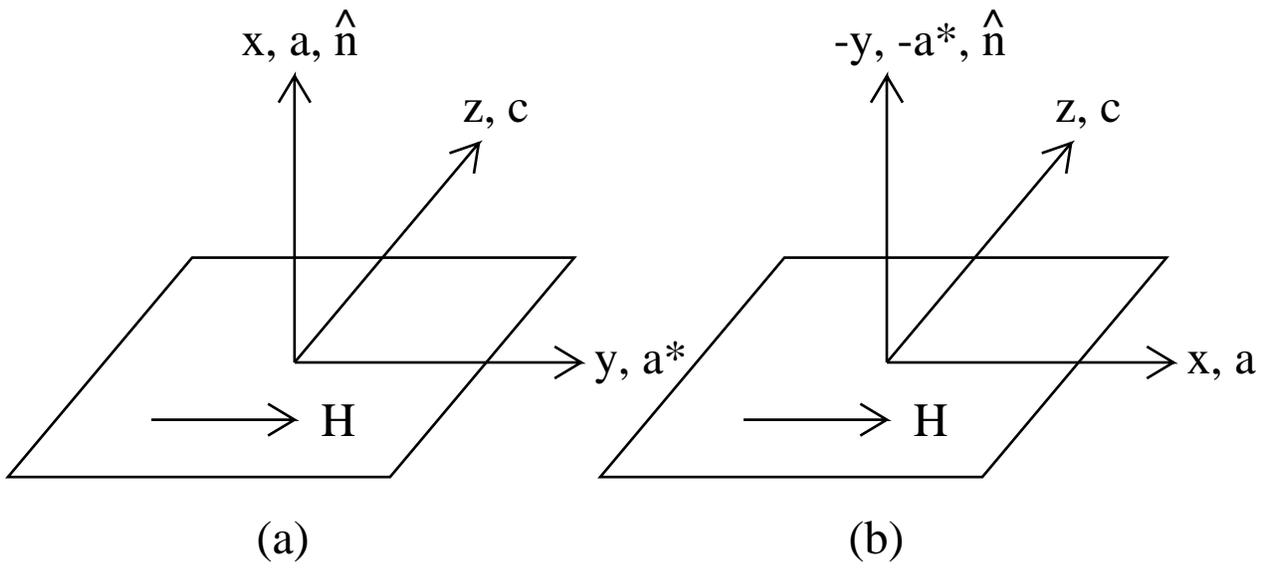
\begin{figure}
\caption{High symmetry geometries when ${\bf\hat n}\perp {\bf\hat z}$,
${\bf H}\parallel{\bf\hat z}\times{\bf\hat n}$. $a$, $a^*$, and $c$ are
the crystal axes.}
\end{figure}

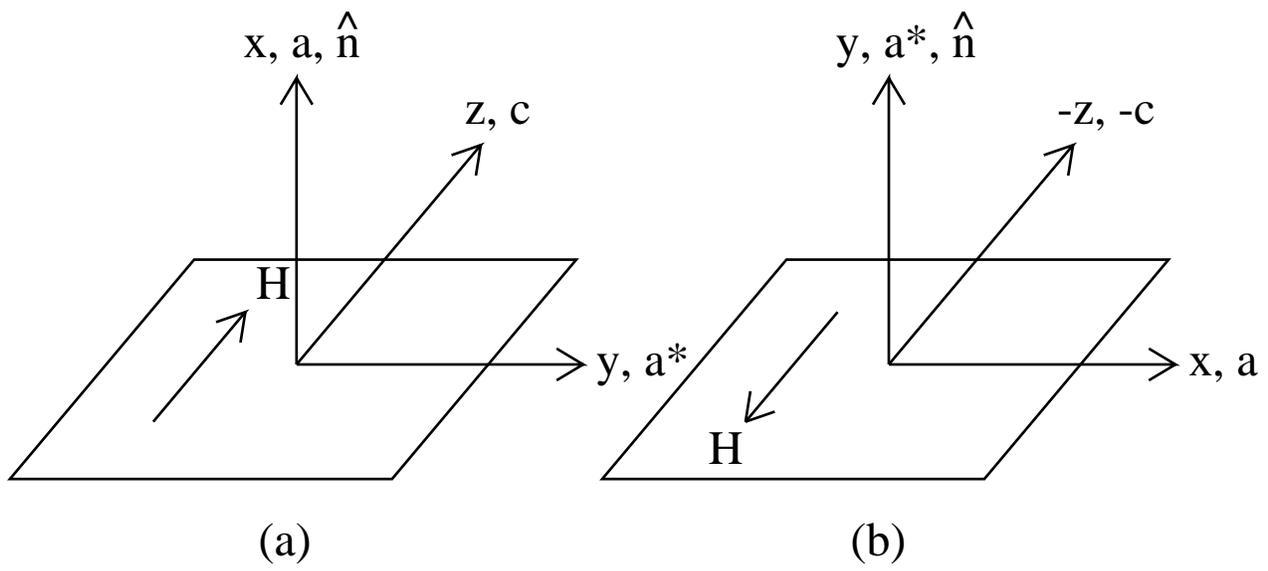
\begin{figure}
\caption{Same as Fig. 4 for ${\bf\hat n}\perp {\bf\hat z}$,
${\bf H}\parallel{\bf\hat z}$.}
\end{figure}

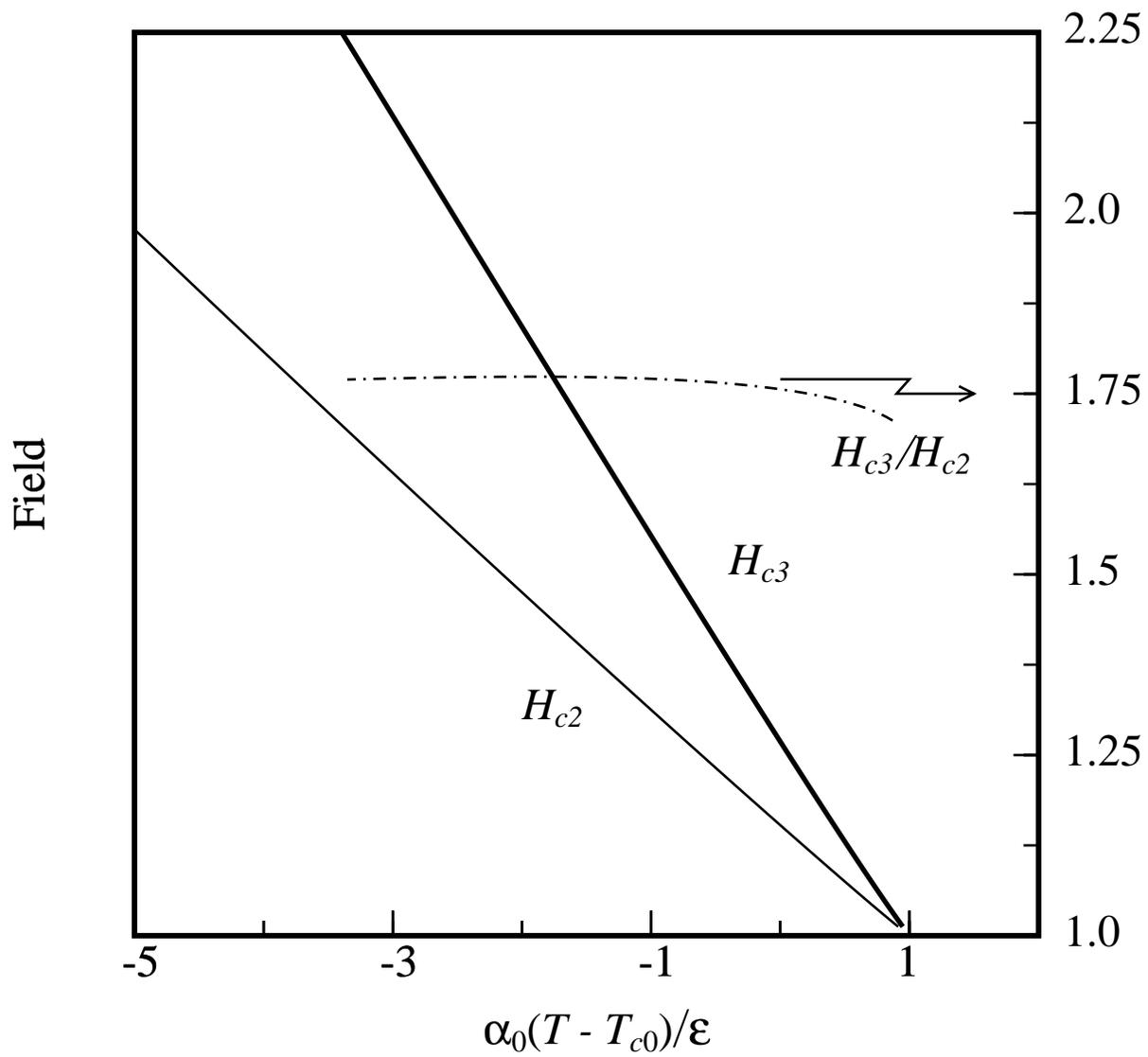
\begin{figure}
\caption{Behavior of $H_{c3}(T)$ in the $E$ models for ${\bf\hat
n}\perp {\bf\hat z}$, ${\bf H}\parallel{\bf\hat z}$, when the higher
$T_c$ component supports surface superconductivity. The dot-dashed line
is the ratio $H_{c3}/H_{c2}$. The curves shown are calculated for
$u=0.46$, $v=0.1$.}
\end{figure}

\begin{table}
\caption{Boundary conditions for various candidate order parameters for
UPt$_3$. The last three columns list the pair of quantities that must
vanish at the surface for given surface normal $\bf\hat n$. Note that
(i) a prime denotes a normal derivative, (ii) the axes $\bf\hat x$,
$\bf\hat y$, and $\bf\hat z$ are fixed along the crystal symmetry axes
$a$, $a^*$, and $c$, respectively.}
\begin{tabular}{ccccc}
 & & & Quantities that vanish for \\
Representation & $\psi_r$ or $\vec \psi_r$ & ${\bf\hat n} =
  {\bf\hat z}$  & ${\bf\hat n} = {\bf\hat x}$ & ${\bf\hat n} =
  {\bf\hat y}$ \\
\tableline
$E_{1g}$ & $(k_zk_x,k_zk_y)$ & $(\eta_1, \eta_2)$ &
  $(\eta_1, \eta_2')$ & $(\eta_1',\eta_2)$ \\
$E_{2g}$ & $(k_x^2 - k_y^2,2k_xk_y)$ & $(\eta_1', \eta_2')$ &
  $(\eta_1', \eta_2)$ & $(\eta_1',\eta_2)$ \\
$E_{1u}$ & $(k_x, k_y)\hat z_s$ & $(\eta_1', \eta_2')$ &
  $(\eta_1, \eta_2')$ & $(\eta_1',\eta_2)$ \\
$E_{2u}^{(1)}$ & $(k_x\hat x_s-k_y\hat y_s, k_x\hat y_s+k_y\hat x_s)$ &
  $(\eta_1', \eta_2')$ & Mixed & Mixed \\
$E_{2u}^{(2)}$ & $(k_x^2-k_y^2, 2k_xk_y)k_z\hat z_s$ &
  $(\eta_1, \eta_2)$ & $(\eta_1', \eta_2)$ & $(\eta_1',\eta_2)$ \\
$A_{1g}\oplus B_{2g}$ & $(1,k_x^3k_z-3k_xk_y^2k_z)$ &
  $(\eta_1', \eta_2)$ & $(\eta_1', \eta_2)$ & $(\eta_1',\eta_2')$ \\
$A_{1u}\oplus B_{2u}$ & $(k_z, 3k_x^2k_y-k_y^3)\hat z_s$ &
  $(\eta_1, \eta_2')$ & $(\eta_1', \eta_2')$ & $(\eta_1',\eta_2)$ \\
$A_{1u}\oplus B_{1u}$ & $(k_z, 3k_xk_y^2-k_x^3)\hat z_s$ &
  $(\eta_1, \eta_2')$ & $(\eta_1', \eta_2)$ & $(\eta_1',\eta_2')$
\end{tabular}
\end{table}

\begin{table}
\caption{Constraints on the transition temperatures, or the sign of the
coupling to the symmetry breaking field, for the order parameters
listed in Table I, obtained by requiring agreement with the $H_{c3}$
data of Keller {\it et al.} for ${\bf H}\parallel {\bf\hat a}^*$,
${\bf\hat n}\parallel{\bf\hat a}$.}
\begin{tabular}{cc}
Order parameter & Constraint \\
\tableline
$A_{1g}\oplus B_{2g}$ & $T_a > T_b$ \\
$A_{1u}\oplus B_{2u}$ & No agreement \\
$A_{1u}\oplus B_{1u}$ & $T_a > T_b$ \\
$E_{1g}$ or $E_{1u}$ & $\zeta < 0$ \\
$E_{2g}$ or $E_{2u}^{(2)}$ & $\zeta>0$
\end{tabular}
\end{table}

\begin{table}
\caption{Expected $H_{c3}$ vs. $T$ behavior for the principal
geometries. The order parameters for each case are as in Table I, and
the signs of ($T_a - T_b$) and $\zeta$ are as in Table II. A blank
denotes a case that we have not studied. `No SSC' stands for no surface
superconductivity.}
\begin{tabular}{cccccc}
Order parameter & ${\bf\hat n} = {\bf\hat c}$ & ${\bf\hat n} =
 {\bf\hat a}$ & ${\bf\hat n} = {\bf\hat a}^*$ & ${\bf\hat n} =
 {\bf\hat a}$ & ${\bf\hat n} = {\bf\hat a^*}$ \\
& ${\bf H}\perp {\bf\hat c}$ &  ${\bf H}\parallel {\bf\hat a}^*$ &
 ${\bf H}\parallel {\bf\hat a}$ &  ${\bf H}\parallel {\bf\hat c}$ &
 ${\bf H}\parallel{\bf\hat c}$ \\
\tableline
$E_{1g}$ & No SSC & Fig. 1 & Fig. 1 & Fig. 6$^a$ & Fig. 6 \\
$E_{2g}$ & Fig. 3 & Fig. 1 & Fig. 2 & Fig. 6$^a$ & Fig. 6$^a$ \\
$E_{1u}$ & Fig. 3 & Fig. 1 & Fig. 1 & Fig. 6$^a$ & Fig. 6 \\
$E_{2u}^{(1)}$ & Fig. 3 & & & &  \\
$E_{2u}^{(2)}$ & No SSC & Fig. 1 & Fig. 2 & Fig. 6$^a$ & Fig. 6$^a$ \\
$A_{1g}\oplus B_{2g}$ & Fig. 1 & Fig. 1 & Fig. 3 & Fig. 1 & Fig. 3 \\
$A_{1u}\oplus B_{2u}$ & Fig. 2$^b$ & Fig. 3 & Fig. 1$^c$ & Fig. 3 &
   Fig. 1$^c$ \\
$A_{1u}\oplus B_{1u}$ & Fig. 2 & Fig. 1 & Fig. 3 & Fig. 1 & Fig. 3
\end{tabular}
a. Assuming that the surface reorients ${\bf M}^\dagger$. Otherwise no
surface superconductivity is expected.

b. Assuming $T_a > T_b$. Otherwise Fig. 1 applies.

c. Assuming $T_a > T_b$. Otherwise Fig. 2 applies.

\end{table}

\begin{references}
\bibitem{refb} In the other subclass the parameters have opposite
parity and are otherwise unrestricted.
\bibitem{refc} If the parity is odd, we replace the `g' suffix by `u'
everywhere.
\bibitem{refgc} This failure arises from the presence of gradient terms
in the free energy that couple the two order parameter components, and
thus lead to level repulsion in the $H_{c2}$ eigenvalue problem.
Barring the weak hexagonal Fermi-surface anisotropy argument
surrounding Eq. (6.6) of Ref. \cite{refGCprb}, this problem is also
present in the $A_1$-$A_2$ and $B_1$-$B_2$ models examined in Ref.
\cite{refMin}, which admit a term $i H_z(\eta_1^*\eta_2 -
\eta_1\eta_2^*)$ in the free energy, where $\eta_1$ and $\eta_2$ are
the two order parameter components.
\bibitem{refd} We do not consider paramagnetic corrections such as
those embodied in Eq.~(19) of Ref.\cite{refSauls}.
\bibitem{reff} The general gap function is a linear combination of
`crystal harmonics' and need not have definite orbital angular
momentum. The examples in Table I suggest otherwise and may be too
simplistic in this regard.
\bibitem{refj} We obtain the same answers using Weber functions for
 $\eta_x$ and $\eta_y$.
\bibitem{refkmix} We have not investigated the onset of $k$-mixing as
$u$ is changed.
\end{references}
\end{document}